\documentclass{mem}
\usepackage{natbib}\usepackage{txfonts}\usepackage{balance}
\usepackage{graphicx}
\usepackage[a4paper,breaklinks,dvipdfm]{hyperref}
\idline{75}{282}
\begin{document}
\def\teff{$T\rm_{eff }$}
\def\kms{$\mathrm {km s}^{-1}$}
\title{
The Role of AGB Stars Feedback in Sustaining Galaxy Evolution
}
\subtitle{}
\author{
Atefeh Javadi\inst{1} 
, Jacco Th. van Loon\inst{2}
\and Habib  Khosroshahi \inst{1}
          }
\institute{School of Astronomy, Institute for Research in Fundamental Sciences
      (IPM), P.O.\ Box 19395-5531, Tehran, Iran, \email{atefeh@ipm.ir}
\and
Astrophysics Group, Lennard-Jones Laboratories, Keele University,
     ST5 5BG, UK  
}
\authorrunning{Javadi}
\titlerunning{M\,33 monitoring}
\abstract{
We have conducted a near-infrared monitoring campaign at the UK InfraRed Telescope,
of the Local Group spiral galaxy M33. The main aim was to identify stars in 
the very final stage of their evolution, and for which the luminosity is more
directly related to the birth mass than the more numerous less--evolved giant
stars that continue to increase in luminosity. In first instance, only the central
square kiloparsec were monitored and analysed, with the UIST camera. Photometry
was obtained for 18,398 stars; of these 812 stars were found to be variable, most
of which are asymptotic giant branch (AGB) stars. We constructed the birth mass 
function and hence derived the star formation history.
These stars are also important dust factories. We measure their dust
production rates from a combination of our data with Spitzer Space Telescope 
mid-IR photometry. The mass loss rates are seen to increase with increasing
strength of pulsation and with increasing bolometric luminosity.
We construct a 2D map of the mass return rate,
showing a radial decline but also local enhancements due to the concentration of 
red supergiants. We conclude
that star formation in the central region of M33 can only be sustained if gas is accreted from further
out in the disc or from circum--galactic regions. By using data of wide-field camera (WFCAM), the campaign 
was expanded to cover two orders of magnitude larger area, comprising the disc
of M33 and its spiral arms. Photometry was obtained for 403,734 stars; of these
4643 stars were found to be variable. We here present the star formation history across 
the disc of M\,33. 
\keywords{stars: evolution --
stars: luminosity function, mass function --
stars: mass-loss --
stars: oscillations --
galaxies: individual: M\,33 --
galaxies: stellar content }
}

\maketitle{}

\section{Introduction}

Galactic evolution is driven at the end--points of stellar evolution, where
copious mass loss returns chemically--enriched and sometimes dusty matter back
to the interstellar medium (ISM); the stellar winds of evolved stars and the
violent deaths of the most massive stars also inject energy and momentum into
the ISM, generating turbulence and galactic fountains when superbubbles pop as
they reach the ``surface'' of the galactic disc. The evolved stars are also
excellent tracers, not just of the feedback processes, but also of the
underlying populations, that were formed from millions to billions of years
prior to their appearance. The evolved phases of evolution generally represent
the most luminous, and often the coolest, making evolved stars brilliant
beacons at IR wavelengths. The final
stages of stellar evolution of stars with main-sequence masses up to $M\sim30$
M$_\odot$ -- Asymptotic Giant Branch (AGB) stars and red supergiants -- are
characterised by strong radial pulsations of the cool atmospheric layers,
rendering them identifiable as long-period variables (LPVs) in photometric
monitoring campaigns spanning months to years (e.g., Whitelock et al.\ 1991).

M\,33 is the nearest spiral galaxy besides the Andromeda galaxy, and seen under
a more favorable angle. This makes M\,33 ideal to study the structure and
evolution of a spiral galaxy. We will thus learn how our own galaxy 
the Milky Way formed and evolved, which is difficult to do directly 
due to our position within its dusty disc. 

Our methodology comprises three stages: 

[1] find stars that vary in brightness with large amplitude
(about a magnitude) and long period (months to years), 
and identify them by their colours and luminosity as cool
giant stars at the endpoints of their evolution;

[2] use the fact that these stars no longer evolve in brightness, to 
uniquely relate their brightness to their birth mass, and use the birth mass 
distribution to construct the star formation history (SFH); 

[3] measure the excess infrared emission from dust produced by these stars, 
to quantify the amount of matter they return to the interstellar medium in M\,33.

\section{Observations}

For observations we used three of UKIRT's imagers:
UIST, UFTI and WFCAM. UIST and UFTI  cover
the central part ($\sim$ 1kpc$^2$ ) while WFCAM covers a
much larger part of M 33 (15 kpc $\times$ 15 kpc). The combined,
square--degree mosaic of M 33 in the K--band is
shown in Fig.\ 1. The square--kpc central area is represented with a box.

\begin{figure*}[t!]
\resizebox{\hsize}{!}{\includegraphics[clip=true]{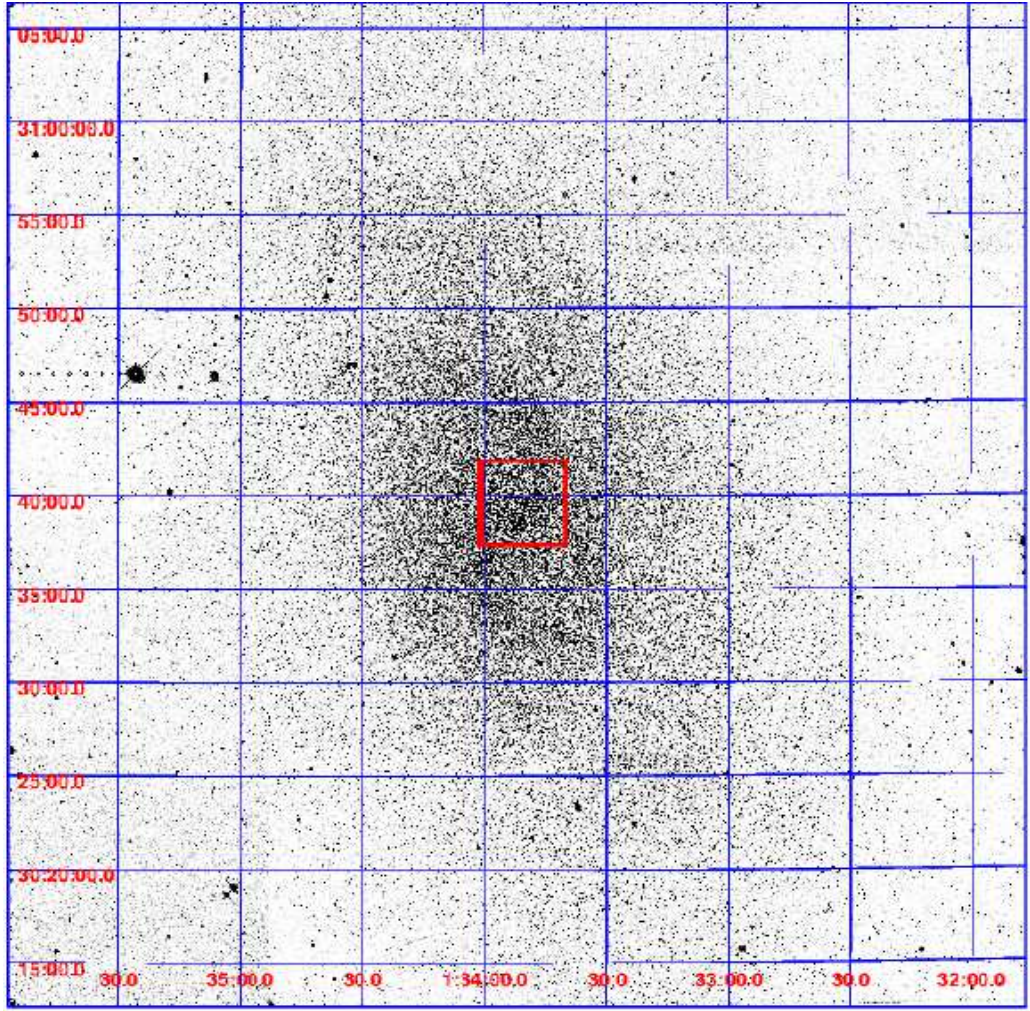}}
\caption{\footnotesize
Combined WFCAM K--band mosaic of M\,33. The square--kpc central area is represented with a box.
}
\end{figure*}

The survey and identification of variable stars are 
described in detail in Javadi et al. (2011, paper I) and
Javadi et al. (2015, paper IV); 812 variable stars were
identified in the central square kiloparsec of M 33 by using
multi--epoch UIST data and 4643 variable stars were
found across the galactic disc of M 33 by using multi--epoch data from WFCAM (Fig.\ 2).

\begin{figure}[]
\resizebox{\hsize}{!}{\includegraphics[clip=true]{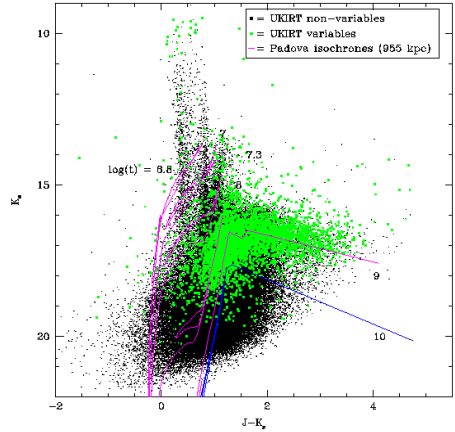}}
\caption{Colour--magnitude diagram of K$_s$  versus $(J-K_{\rm s})$, with WFCAM variable stars in green.
Overplotted are isochrones from Marigo et al.\ (2008) for solar metallicity
and a distance modulus of $\mu=24.9$ mag.
\footnotesize
}
\end{figure}

\section{The Star Formation History in M\,33}

In paper II, we have developed a novel way to derive the star formation
 history by using large--amplitude variable stars which are identified 
 in our IR monitoring programme. We used the fact that these variables 
 have reached the very final stages of their evolution, and their brightness can thus 
 be transformed into their mass at birth by employing theoretical evolutionary 
 tracks or  isochrones. The LPVs are located at the cool end of each 
 of the isochrones and this is confirmed by inspection of isochrones.
 Therefore, we simply could construct a link between the observed K--band  
 magnitude and theoretical models to estimate the birth mass of LPVs.
  This is done in paper II for four different metallicities from 
  super--solar metallicity suitable for massive elliptical galaxies and 
  stellar populations in the bulge of massive spiral galaxies such as the Milky Way to 
  sub--solar values applicable the Large and Small Magellanic Clouds (Rezaeikh et al.\ 2014). 
  As discussed in  paper II, the central region of M\,33 has approximately solar 
  metallicity, so we could adopt Z=0.015.
   The outer regions of M\,33 have lower metallicities  than the center. Based 
   on discussions in paper IV, it seems Z=0.008 agrees well with colour--magnitude 
   diagrams (CMDs), so  we adopt Z=0.008 for the disc and spiral arms of M\,33. 
   The star formation history is estimated by:
   \begin{equation}
\xi(t)=\frac{{\rm d}n^\prime(t)}{\delta t}\ \frac{\int_{\rm min}^{\rm
max}f_{\rm IMF}(m)m\,{\rm d}m}{\int_{m(t)}^{m(t+{\rm d}t)}f_{\rm IMF}(m)\,{\rm
d}m}.
\end{equation}
Where n$^\prime$ is the number of variables that we have 
identified, $f_{IMF}$ is the initial mass function describing the
relative contribution to star formation by stars of different
mass and $\delta$t is the duration of variability which
these stars display strong radial pulsation.

The SFH of the central square kpc of M\,33 using WFCAM data is shown 
in Fig.\ 3, overlayed with red points  by the SFH which was derived by using UIST data.
Two main epochs of star formation are obvious; a major epoch of formation 
$\approx$ 4--5 Gyr ago ($\log t$=9.6--9.7) peaking around 4 Gyr ago at a level 
about 2.5 times as high as during the subsequent couple of Gyr.
A second epoch of star formation is seen to  occur from 
200 Myr--300 Myr ago ($\log t$=8.3--8.5), with a rate around 
1.5 times higher than the first peak at 4 Gyr ago. Since then the rate of star formation 
is decreasing.

\begin{figure}[]
\resizebox{\hsize}{!}{\includegraphics[width=400mm]{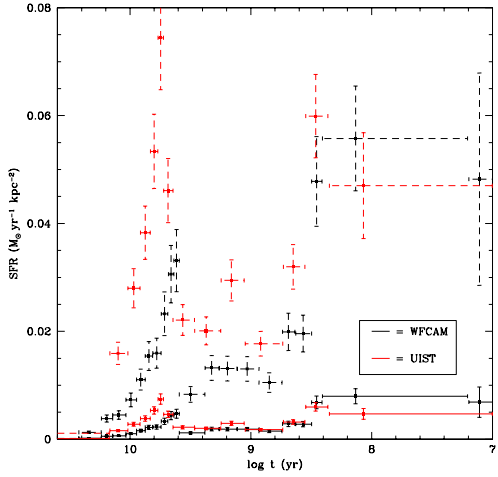}}
\caption{The star formation history in the central square kpc 
of M\,33 derived from pulsating AGB stars and red--supergiants;(black) 
WFCAM data is used, (red) UIST data is used. (dashed line) Correction factor 
has been applied to star formation rates.
\footnotesize
}
\end{figure}

The total area which is covered by WFCAM observations is 
almost 15 kpc $\times$ 15 kpc, which covers the 
pseudo bulge, disc and spiral arms of the  M\,33 galaxy.
The SFH across the disc of M\,33 is shown in Fig.\ 4. Here again we see 
two epochs of enhanced star formation. The old epoch of star formation 
is seen at $\approx$ 3--6 Gyr ago ($\log t$=9.5--9.8) peaking around 5 Gyr ago ($\log t$=9.7) 
with a level almost 2 times as high as during the subsequent few Gyr. Barker et al. \ 2011 have 
shown that the outer disc of M\,33 had a major epoch of star formation $\sim$ 2--4 Gyr ago, or at 
z $\sim$ 0.2--0.4. The 
peak at 4 Gyr ago resembles the peak that they have reported by using CMD fitting to the 
two fields in the outer disc of M\,33 observed with the Hubble Space Telescope Advanced Camera.
The recent epoch of enhanced star formation 
is occurring $\approx$ 200--300 Myr ago ($\log t$=8.3--8.5), reaching a level 
of almost 4 times larger than the peak at the first epoch of star formation. This second 
peak is very strong compared with what we saw for the center 
of M\,33.

\begin{figure}[]
\resizebox{\hsize}{!}{\includegraphics[width=400mm]{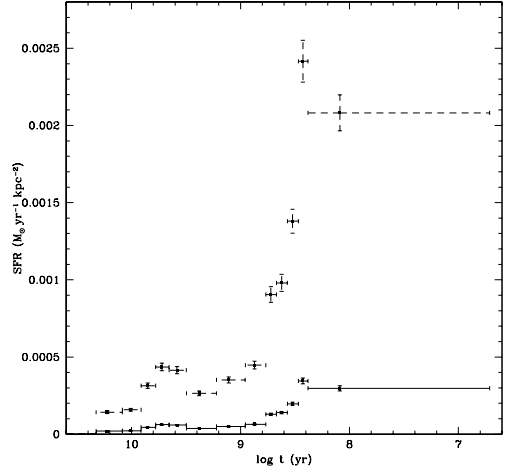}}
\caption{The star formation history across the disc of M\,33, before the
correction on pulsation duration was applied
(solid line) and after the correction 
on pulsation duration was applied (dashed line).
\footnotesize
}
\end{figure}

The star formation history variations versus distance from the center (r) 
is shown in Fig.\ 5. Each radius bin contains the same number of variable stars.  This figure 
reveals two important results immediately; Firstly, the old epoch of star formation 
in M\,33 is weakening and then disappears as we go further out from the center. 
The star formation starts and then gradually increases until 
t= 200 Myr ($\log t$=8.3) and then decreases. Secondly, the very 
recent epoch of star formation is seen in all bins across 
the disc with same strength. 
 It clearly shows the inside--out formation of M\,33.

\begin{figure}[]
\resizebox{\hsize}{!}{\includegraphics[width=400mm]{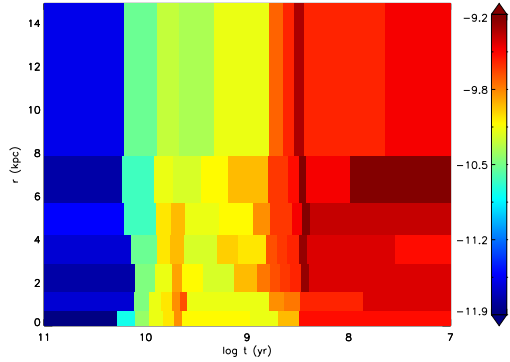}}
\caption{The star formation history  of M\,33 at different radius bins. The star 
formation rates are normalized to the total stellar mass per square kiloparsec 
in each radius bins.The colour bar values are in 
logarithmic scale.
\footnotesize
}
\end{figure}

\section{Mass Loss Rates in the Central Square Kiloparsec of M\,33}

We derive the mass--loss rates of the red giant variables in two steps;
First we model the spectral energy distribution (SED) of near--
IR variables for which we have Spritzer Space Telescope mid--IR data (Fig. 6) and
then we use these results to construct a relation between
the dust optical depth and bolometric corrections on the
one hand, and near--IR colours on the other. Then we
apply those relations  to the
red giant variables for which no mid-IR counterpart was
identified, to derive their mass-loss rates too (Fig. 7).
The total mass return from UKIRT variables is almost
$\sim$ 0.0055 M$_\odot$ yr$^{-1}$ and carbon grains make up $<$ 23 $\%$
of the present--day dust--mass return so the interstellar
dust is predominantly oxygen--rich. The full 2--D map 
of mass return is shown in Fig.\ 8.

\begin{figure}[]
\resizebox{\hsize}{!}{\includegraphics[clip=true]{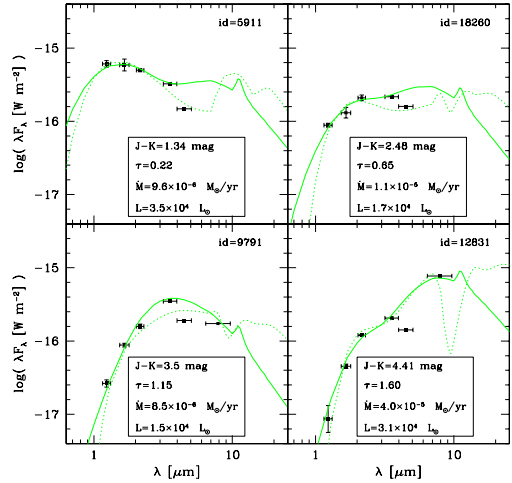}}
\caption{Near-- and mid--IR photometry of examples of carbon
stars in the centre of M 33, affected by various levels of
mass loss. The solid lines are
the best matching SEDs modelled with dusty. The dotted
lines are best matching fits using silicates, for comparison.
\footnotesize
}
\end{figure}

\begin{figure}[]
\resizebox{\hsize}{!}{\includegraphics[clip=true]{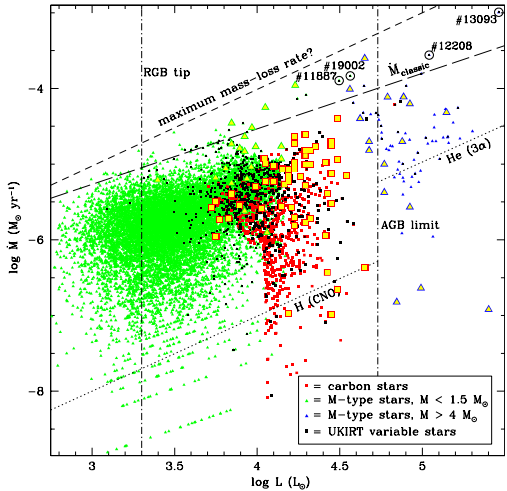}}
\caption{Mass--loss rate vs. luminosity. Massive, luminous
M--type stars are depicted by blue triangles; AGB carbon
stars by red squares; and low--mass M--type stars
by green triangles. Large yellow symbols identify
the stars modelled with dusty; other UKIRT variable stars
are identified by black squares. The most extreme mass 
losing stars are labelled.
\footnotesize
}
\end{figure}
The estimated ISM depletion timescale by Kang et al.
(2012) is 0.3 Gyr. The mass return from evolved
stars would not change the timescale by more than 17
$\%$. If we account for the mass return from supernovae, hot
massive--star winds, luminous blue variable eruptions, et
cetera, the mas return rate increases from $\sim$0.004–-0.005
M$_\odot$ yr$^{-1}$ kpc$^{-2}$ to $\sim$ 0.006 M$_\odot$ yr$^{-1}$ kpc$^{-2}$.
Therefore,
the above conclusion does not change. For sustaining
star formation with the current rate gas must flow into
the central regions of M 33, either through a viscous disc
or via cooling flows from the circum--galactic medium.
\begin{figure*}[t!]
\centerline{\hbox{{\includegraphics[width=60mm]{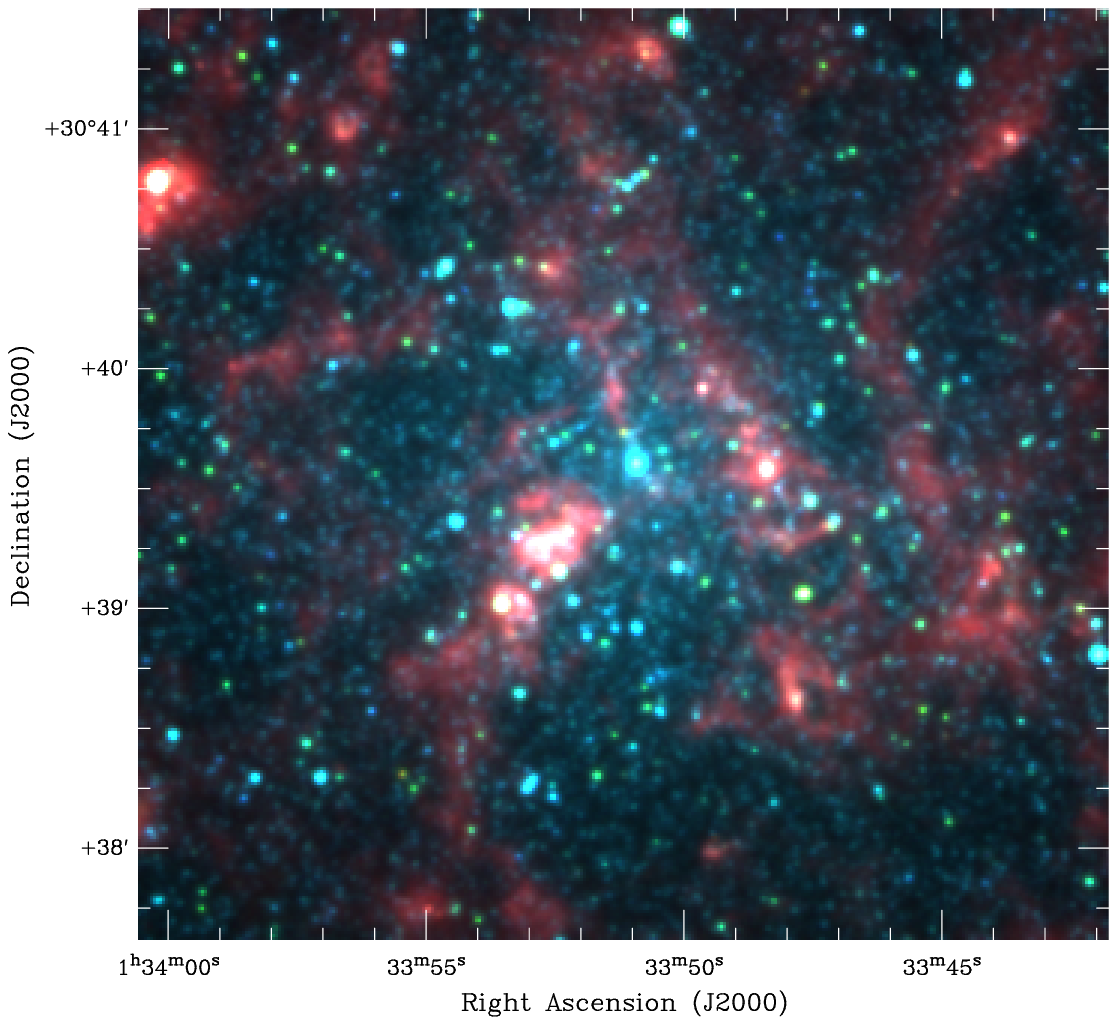}}
{\includegraphics[width=73mm]{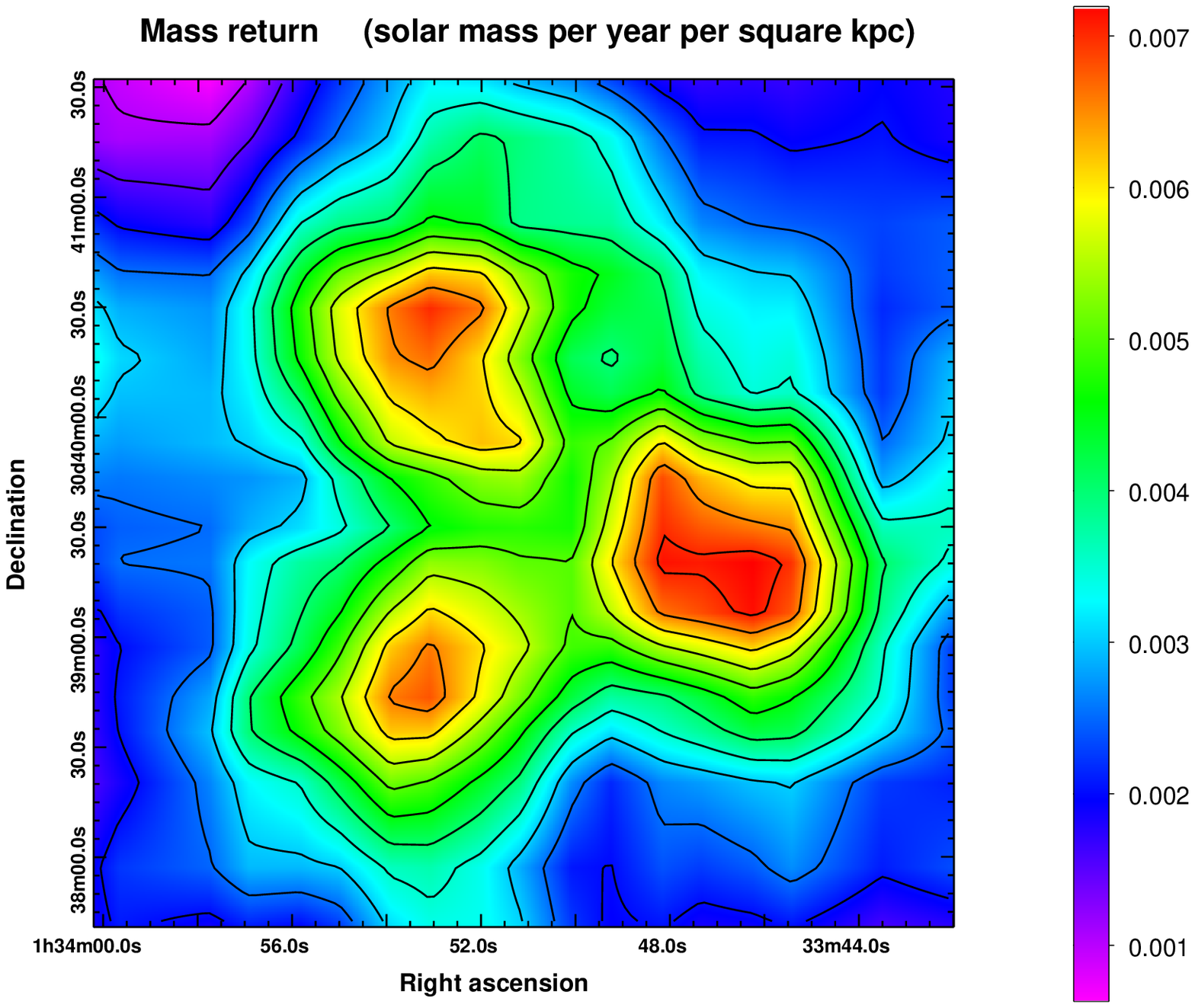}}}}
\caption{\footnotesize
Left: Spitzer composite image of IRAC bands 1, 2 and 4 at respectively 
3.6 $\mu$m (blue), 4.5 $\mu$m (green) and 8 $\mu$m (red); Right: map of
mass-return-rate surface density over the central region of M 33.
}
\end{figure*}

In paper III we calculated the ratio of the mass lost during the pulsation phase to the birth mass 
and we realized that the pulsation duration is over--estimated by theoretical models. 
If the pulsation duration was over-estimated, then the
star formation rate would have been under-estimated. Comparison
 of recent star formation rate derived from our method with the  
 values derived by other methods shows that star formation rate 
 should be corrected by the factor of 10 in the central regions and 7 for the disc.
 In Fig.\ 3 and Fig.\ 4
  the correction factor of 10 for the center and 7 for the disc is applied to the star formation rates.
\section{Conclusions}
The photometric catalogues of Javadi et al. (2011) and Javadi et al. (2015) were used 
to reconstruct the star formation history 
across the galactic disc of M\,33. The numbers and luminosities of the 
pulsating AGB stars and red supergiants were converted to star formation rate 
as a function of look-back time, using Padova stellar evolution models (Marigo et al. 2008). 
The pulsating red giant and supergiant stars were used to map the dust production 
in the central square kiloparsec of M\,33. 

We are currently working on WFCAM data to estimate the mass loss rates 
of evolved stars across the galactic disc of M\,33 and to establish a link 
between the dust return and the formation of stars within the prominent 
spiral arm pattern.
\begin{acknowledgements}
We acknowledge support from the "Royal Society International Exchanges grant IE130487".
JvL acknowledges the award of a Santander Bank Research Bursary to attend this meeting.
\end{acknowledgements}

\bibliographystyle{aa}

\end{document}